\documentstyle[12pt]{book}

\title{On the use of algebraic programming in the general relativity
\footnote{Talk given at The Albert Einstein Institut, Max Planck Institut
f\" ur Gravitationstheorie, Golm, Germany, september 2000  }} 
\author{Dumitru N. Vulcanov \\
The West University of Timi\c soara\\
Theoretical and Computational Physics Department\\
B-dul V. P\^ arvan, Nr. 4, RO-1900 Timi\c soara, Rom\^ ania}

\date{}

\begin{document}

\maketitle

\tableofcontents

\chapter*{Preview}
\addcontentsline{toc}{chapter}{Preview}

This is a review devoted to some results of  Algebraic Programming
(Computer Algebra - CA) used in treating several problems of general 
relativity, based mainly on already published articles (see the references).

The first chapter presents  procedures in REDUCE language using 
EXCALC package for algebraic programming of the Hamiltonian
formulation  of general relativity (ADM formalism). The procedures
calculate the dynamic and the constraint equations and, in addition, we have
extended the obtained procedures in order to perform a complete ADM
reductional procedure. Several versions of the procedures were realized for
the canonical treatment of  pure gravity, gravity in interaction with
material fields, inflationary models (based on a scalar field non-minimally
coupled with gravity) and theories with higher order lagrangians.

The second chapter is devoted to the same problem as the first one, but 
now we use Maple V platform. In this purpose
we present  Maple procedures using the GrTensorII package adapted for 
the study of the canonical version of the general
relativity based on the ADM formalism.

Last chapter presents CA procedures and routines  applied
to the Dirac field on curved spacetimes. The main part of the procedures is 
devoted to the construction of Pauli and Dirac matrices algebra on an 
anholonomic orthonormal reference frame. Then we shall present how these 
algebraic programming sequences are used in the study of the Dirac equation on 
curved spacetimes and noninertial reference frames. 
A comparative review of such procedures obtained for the
two mentioned CA platforms (REDUCE + EXCALC and MAPLE + GRTensorII) 
is presented. Applications for the calculus of Dirac equation on specific 
examples of spacetimes with or without torsion and for the study of the 
non-relativistic approximation of the Dirac field are pointed out, including 
the search for inertial effects.

\chapter{REDUCE programming and the Hamiltonian version of general relativity}
\chaptermark{REDUCE programming and ...}

\section{Introduction}

The Hamiltonian formalism of the general relativity 
\index{Hamiltonian!formalism} is based on the (3+1)-dimensional split of
space-time - (\cite{1}, \cite{2}, \cite{7}. For instance, the Hamiltonian 
formulation and
quantization of some homogeneous and inhomogeneous cosmological models \cite
{3} needs the complete expressions of the super-Hamiltonian 
\index{Hamiltonian!super-} ${\cal {H}}$ and super-momentum 
\index{momentum!super-}${\cal {H}}^{i}$ (see below) and the dynamic
equations in term of the canonical variables. For this purpose, concerning
the great volume of calculations and the division of the calculus in
distinct steps it is possible to create algebraic procedures which transpose
the specific manipulations of the canonical formulation of general
relativity in computer language.

There are some early results in this direction \cite{4}, obtained in old
versions of the REDUCE language, without EXCALC \index{EXCALC} package.
Traditionally, the programs calculate only the dynamic equations and the
constraint equations \index{constraint equations} after introducing the
canonical variables and their canonically conjugate momenta and do not
perform the complete reductional formalism in order to point out the true
dynamic content of the treated model.

In the present chapter we shall summarize recent results obtained
processing some space-time models with several new procedures realized with
EXCALC package (in REDUCE). In fact, here we have a
review of a series of articles (\cite{4} - \cite{6}, \cite{7}) 
containing the details
of our procedures and the obtained results for some concrete space-time
models.

We have extended the programs in order to realize a complete reductional
procedure \index{reductional procedure} (solving the constraint equations,
changing of variables, reduction of dynamic variables, etc.) for some
space-time models.

We shall present also the version of our procedures for the Hamiltonian
treatment of some inflationary models \index{inflationary models} (based on
a scalar field non-minimally coupled with gravity).

\section{The local form of the canonical formalism of gravity}
\sectionmark{The local form ...}

Here we shall use the specific notations for the ADM formalism 
\index{ADM formalism} \cite{1},\cite{2}; for example latin indices will run
from 1 to 3 and greek indices from 0 to 3. The starting point of the
canonical formulation of the general relativity is the (3+1)-dimensional
split of the space-time produced by the split of the metric tensor : 
\begin{equation}\label{1} 
^{(4)}g_{\alpha\beta} = \left ( 
\begin{array}{ccc}
^{(4)}g_{oo} & ~~ & ^{(4)}g_{oj} \\ 
~~ & ~~ & ~~ \\ 
^{(4)}g_{io} & ~~ & ^{(4)}g_{ij} 
\end{array}
\right ) = \left ( 
\begin{array}{ccc}
N_{k}N^{k}-N^{2} & ~~ & N_{j} \\ 
~~ & ~~ & ~~ \\ 
N_{i} & ~~ & g_{ij} 
\end{array}
\right ) 
\end{equation}
where $g_{ij}$ is the riemannian metric tensor of the three-dimensional
spacelike hypersurfaces at t = const. which realize the spacetime foliation.
Here $N$ is the "lapse" \index{lapse} function and $N^{i}$ are the
components of the "shift" vector \cite{2}.

The Einstein vacuum field equations now are (denoting by "$\cdot$" the time
derivatives) : 
\begin{equation}\label{2} 
\dot{g}_{ij} = 2Ng^{-1/2}[\pi_{ij}-\frac{1}{2}g_{ij}{\pi^k}%
_k]+N_{i/j}+N_{j/i} 
\end{equation}
\begin{eqnarray}
\dot{\pi}^{ij} = -Ng^{1/2}[R^{ij}-\frac{1}{2}g^{ij}R]+
       \frac{1}{2}Ng^{-1/2}g^{ij}[\pi^{kl}\pi_{kl}-\frac{1}{2}({\pi^k}_k)^{2}] \nonumber
\end{eqnarray}
\begin{eqnarray}
~~~~~~~~~~-2Ng^{-1/2}[\pi^{im}{\pi^j}_{m}-\frac{1}{2}\pi^{ij}{\pi^k}_k] +
                  g^{1/2}[N^{/ij}-g^{ij}{N^{/m}}_{/m}] \nonumber
\end{eqnarray}
\begin{equation}\label{3}
 ~~~~~~~~+[\pi^{ij}N^{m}]_{/m} - {N^i}_{/m}\pi^{mj} - {N^j}%
_{/m}\pi^{mi} 
\end{equation}
where $\pi^{ij}$ are the components of the momenta canonically conjugate to
the $g_{ij}$'s.

In the above formulas we denoted by "/" the three-dimensional covariant
derivative defined with $g_{ij}$ using the components of the
three-dimensional connection $\Gamma^{i}_{jk}$ and $R_{ij}$ is the
three-dimensional Ricci tensor.

The initial data on the t = const hypersurface are not independent because
they must satisfy the constraint equations, which complete the Einstein
equations 
\begin{equation}\label{6} 
{\cal {H}}= -\sqrt{g} \lbrace R+g^{-1}[\frac{1}{2}({\pi^k}%
_{k})^{2}-\pi^{ij}\pi_{ij}]\rbrace =0 
\end{equation}
\begin{equation}\label{7} 
{\cal {H}}^i=-2{\pi^{ij}}_{/j} =0 
\end{equation}
where ${\cal {H}}$ is the super-Hamiltonian, ${\cal {H}}^{i}$ the
super-momentum and $g$ is the determinant of the three-dimensional metric
tensor $g_{ij}$. The action functional in Hamiltonian form for a vacuum
space -time can thus be written as (\cite{1},\cite{2}) 
\begin{equation}\label{8} 
S=\int{dt \int{\ (\pi^{ij} \dot{g}_{ij} - N{\cal {H}} - N_{i}{\cal 
{H}}^{i})}} \omega^{1}\omega^{2}\omega^{3} 
\end{equation}
where the $\omega^i$'s are the basis one-forms.

In Dirac's version of the Hamiltonian formalism \cite{2} the constraints
equations (\ref{6}) and (\ref{7}) are not solved and no coordinate condition
is imposed; The Dirac Hamiltonian is then given by \index{Hamiltonian!Dirac} 
\begin{equation}\label{9} 
H = \int{(N{\cal {H}}+N_i{\cal {H}}^i)}\omega^{1}\omega^{2}%
\omega^{3} 
\end{equation}

Thus the dynamic equations (\ref{2}) and (\ref{3}) are obtained by
differentiating $H$ with respect to the canonical conjugate pair of
variables $(\pi^{ij},g_{km})$.

When interaction with a scalar field $\chi$ is considered, the free
gravitational action must be completed with several new terms. The
lagrangian of the massless scalar field is \cite{10} $L_{sc} =
-(-^{(4)}g)^{1/2}~^{(4)}g^{\mu \nu} \chi_{,\mu} \chi_{,\nu}$ and defining
the scalar field momentum as $\pi_{\chi} = (\partial L_{sc}/\partial
\chi_{,t})$ we must add to the total gravitational action : 
\begin{equation}\label{10} 
S_{sc} = \int dt \int (\pi_{\chi} \dot{\chi} - N(\frac{
\pi_{\chi}^2 }{4\sqrt{g}} + \sqrt{g} g^{ij} \chi_{,i} \chi_{,j} ) - N_i
(\pi_{\chi}g^{ij} \chi_{,j} ) ) \omega^{1}\omega^{2}\omega{3} 
\end{equation}

It is obvious from the above formula that the terms in brackets must be
added to the gravitational super-Hamiltonian and to the super-momentum. Thus
we can obtain the new form of the constraint equations and Dirac
Hamiltonian. With the last one it can obtain the dynamic equations for the
scalar field variables.

When interaction with a perfect fluid (consisting of pressureless dust) is
present, to the total gravitational action we can add \cite{11} : 
\begin{equation}\label{11} 
S_{d} = \int dt \int (\pi_{\phi} \dot{\phi} - N \pi_{\phi} 
\left [ 1+ g^{ij} \phi_{,i} \phi_{,j} \right ]^{\frac{1}{2}}- N_i g^{ij}
\phi_{,j} ) \omega^1 \omega^2 \omega^3 
\end{equation}
where $\phi$ is the scalar field describing the dust.

In the last few years, inflationary models \index{inflationary models}
became very popular not only in cosmology but also in the theory of general
relativity and gravitation. The majority of the mentioned models are based
on the interaction of gravity with a scalar field \cite{15} -- \cite{17}, 
\cite{37}, \cite{38}. We
shall treat a model based on the action functional : 
\begin{equation}\label{12}
S=\int d^4x\sqrt{-^{(4)}g}\left( \Phi ^2~^{(4)}R-\frac
12~^{(4)}g^{\mu \nu }\Phi _{,\mu }\Phi _{,\nu }+\Lambda \Phi ^4\right) 
\end{equation}
where the scalar field $\Phi $ is non-minimally coupled with the gravity. As
an inflationary model the above action functional determines the behavior of
the gravitational constant as being $G=1/16\pi \Phi ^2$ through the dynamics
of the field $\Phi $.

In order to obtain the canonical version of the field theory described by
the action functional (\ref{12}) we must redefine the components of the
three-dimensional momentum canonical conjugate to the $g_{ij}$'s as : 
\begin{equation}\label{13} 
\pi^{ij} = \Phi^2 \sqrt{g} \left ( g^{ij} K - K^{ij} \right ) 
\end{equation}
where $K_{ij}=\frac{1}{2N} \left ( N_{i/j} + N_{j/i} - \dot{g}_{ij} \right )$
is the extrinsic curvature of the three-dimensional hypersurfaces. With $K$
we have denoted the trace of $K_{ij}$, namely $K={K^i}_i$.

Thus, after an appropriate calculation presented in complete details in 
\cite{6} we obtain the complete canonical version of our inflationary model 
(\ref{12}) : 
\begin{equation}\label{14}
S=\int dt~d^3x~\left( \pi ^{ij}\dot g_{ij}+\pi _\Phi \dot \Phi -N%
{\cal {H}}-N_i{\cal {H}}^i\right) 
\end{equation}
where the new super-Hamiltonian ${\cal {H}}$ and super-momentum ${\cal {H}}%
^i $ (and the new constraint equations) are: 
\begin{eqnarray}
{\cal{H}} = - \sqrt{g} \Phi^2 R + \frac{1}{\sqrt{g}} \Phi^{-2}
                  \left ( \pi_{ij}\pi^{ij} - \frac{1}{2} \pi^2 \right ) +
 \frac{\pi_{\Phi}^2}{2\sqrt{g}} - \frac{2}{\sqrt{g}} \Phi^{-1} \pi_{\Phi} \pi +
\nonumber
\end{eqnarray}
\begin{equation}\label{15}
\frac 92\sqrt{g}g^{ij}\Phi _{,i}\Phi _{,j}+4\sqrt{g}g^{ij}\Phi
\Phi _{,kj}-\sqrt{g}\Lambda \Phi ^4=0 
\end{equation}
\begin{equation}\label{16}
{\cal {H}}^i=-2{\pi ^{ij}}_{/j}-4g^{ij}\Phi ^{-1}\Phi _{,j}\pi
+g^{ij}\pi _\Phi \Phi _{,j}=0 
\end{equation}

The dynamical equations are obtained in the usual way. For example, after
defining the Dirac Hamiltonian as : 
\begin{equation}\label{17} 
H_D = \int dt~d^3x~ (N{\cal {H}} + N_i {\cal {H}}^i ) 
\end{equation}
we can set : 
\begin{eqnarray}
\dot{g}_{ij} = \frac{\delta H_D}{\delta \pi^{ij}}\hbox{~~;~~}
\dot{\pi}^{ij} =- \frac{\delta H_D}{\delta g_{ij}}\hbox{~~;~~}\nonumber
\end{eqnarray}

As a result we have : 
\begin{eqnarray} 
\dot{g}_{ij} = N_{i/j} + N_{j/i} + \frac{2}{\sqrt{g}} N \Phi^{-2} \pi_{ij}
                - \frac{1}{\sqrt{g}} N \Phi^{-2} \pi g_{ij} - \nonumber
\end{eqnarray}
\begin{equation}\label{18} 
\frac{2}{\sqrt{g}} N \Phi^{-1} \pi_{\Phi} g_{ij} - \Phi^{-1}
\Phi_{,p} N_k g^{kp} g_{ij} 
\end{equation}
\begin{eqnarray}
\dot{\pi}^{ij} = 
   - N \Phi^{-2} \sqrt{g} \left ( R^{ij} - \frac{1}{2} R g^{ij} \right ) +
    \frac{1}{2\sqrt{g}} N \Phi^{-2} g^{ij} \left ( \pi^{km}\pi_{km} - \frac{1}{2} \pi^2 \right )-
\nonumber
\end{eqnarray}
\begin{eqnarray}
\frac{2N}{\sqrt{g}} \Phi^{-2} \left ( \pi^{im}{\pi^j}_m - \frac{1}{2} \pi \pi^{ij} \right ) -
\sqrt{g} \Phi^{2} \left ( N^{/ij} - N_{,l}N^l g^{ij} \right ) +
 \nonumber
\end{eqnarray}
\begin{eqnarray}
(\pi^{ij}N^m)_{,m} - \pi^{im}N^j_{/m} - \pi^{jm}N^i_{/m} -
     \frac{1}{4\sqrt{g}} N g^{ij} \pi_{\Phi}^2 + 
         \frac{17}{4}\sqrt{g} N g^{kl} g^{ij} \Phi_{,k}\Phi_{,l}
\nonumber
\end{eqnarray}
\begin{equation}\label{19} 
+ \frac{17}{2}\sqrt{g} N \Phi^{,i} \Phi^{,j} - \frac{1}{2} \sqrt{g%
} g^{ij} N \Lambda \Phi^4 + \Phi^{-1} \left ( N^i \Phi^{,j} + N^j \Phi^{,i}
\right ) 
\end{equation}

Of course, in a similar manner we can obtain the dynamic equations for the
scalar field $\Phi$ and his conjugate momenta $\pi_{\Phi}$. We do not give
here the mentioned equations because, in the algebraic procedures we shall
present in the next paragraph we have obtained these equations directly from
the Dirac Hamiltonian $H_D$ by a variational derivative (a facility of
EXCALC package \cite{13}). The complete form of the $\dot{g}_{ij}$'s and $
\dot{\pi}^{ij}$'s are necessary as a consequence of the fact that in
practice we use the components of the above tensors as functions of the true
canonical variables introduced in every specific case of interest.

\section{The REDUCE computer procedures}

We shall briefly present the steps of calculus necessary for developing the
canonical program outlined above. The steps will be the same if the calculus
is realized manually or with the computer, using specific EXCALC procedures.
Several steps are the same if we have pure gravity, gravity coupled with
matter fields or inflationary models. Thus we shall present the next steps
together with examples of computer lines from the pure gravity topic. Here
are the steps :

1) introduction of the basis vectors and coframe forms (with a usual metric
statement in EXCALC \cite{13}) together with the calculus of the structure
coefficients of the Cartan basis considered; it is realized, in the same
time the introduction of the metric $g_{ij}$ components, because, in
EXCALC package the assignment of the coframe forms defines, in the same time
the metric components;

2) introduction of the momentum $\pi^{ij}$ defining the canonical
conjugates of the metric components and their dependence on the spatial
variables defining a symmetric 2-form, named PIU; here is introduced a
program line computing the term $\pi^{ij}\frac{dg_{ij}}{dt}$ from eq. (\ref
{8}) and  verifying if the dynamic part of the action functional is in canonical
form :

\begin{verbatim}
act := piu(j,k)*@(g(-j,-k),t);
\end{verbatim}
if the answer is negative this step must be run again, after a new
definition of the $\pi^{ij}$ components;

3) evaluates the Ricci tensor $R_{ij}$ and the Ricci scalar $R$ of the
metric above defined by direct computation from a RIEMANNCONX form
named OM (a riemannian connection form), declared in the previous lines of
the program 
\begin{verbatim}
ricci(-j,-m) := ee(-m)_|(ee(-k)_|(d om(k,-j) + 
                              om(k,-p)^om(p,-j)));
    scricci := ricci(j,-j);
\end{verbatim}
where $ee(-k)$ are the basis vectors defined together with the coframe;

4) calculates the super-Hamiltonian ${\cal {H}}$ : 

\begin{verbatim}

ha0:=   -sqrt(detm!*)*(scric+(1/detm!*)*

         ((1/2)*( (piu(j,-j))*(piu(k,-k)) )

              -piu(j,k)*piu(-k,-j)  ) );

\end{verbatim}
from equation (\ref{6}) or from equation (\ref{15}) in our inflationary
model (\ref{12}) ;

5) calculates the super-momentum ${\cal {H}}^i$ : 
\begin{verbatim}
pform ha(j)=0; 
ha(j):=
  -2*((ee(-k)_|d piu(j,k))+(ee(-k)_|om(j,-p))*piu(p,k)
                +(ee(-k)_|om(k,-p))*piu(j,p)
                -(ee(-k)_|om(p,-p))*piu(j,k));
\end{verbatim}
with equation (\ref{7}) or eq. (\ref{16}) for the inflationary model;

6) at this stage it is already possible to calculate the temporal
derivatives for the $g_{ij}$'s components (eq. (\ref{2}) or (\ref{18})) : 
\begin{verbatim}
pform n=0,derge(j,k)=0; tvector ni(k); 
fdomain n=n(r,t),ni=ni(r,t);
derge(-j,-k):=  2*n*(1/sqrt(detm!*))*
     ( piu(-j,-k)-(1/2)*g(-j,-k)*(piu(p,-p)) ) +
  (ee(-k)_|d ni(-j)) - (ee(-k)_|om(p,-j))*ni(-p) +
  (ee(-j)_|d ni(-k)) - (ee(-j)_|om(l,-k))*ni(-l);
\end{verbatim}
(where $n$ and $ni$ are the "lapse-shift" functions) and then the evolution
equations of the canonical variables defined in step 1;

7) calculates the second set of temporal derivatives for $\pi^{ij}$, with
eq. (\ref{3}) (or (\ref{19})) and the dynamic equations for momenta defined
in step 2;

All the above mentioned steps are naturally transposed in EXCALC lines, as
we can see from the above examples. Special program lines are devoted to the
calculus of the new terms appearing in the theory, when interaction of
gravity with a matter field is considered. It is calculated the new Dirac
Hamiltonian (obtained adding new terms to the super-Hamiltonian and the
super-momentum specific for the matter field). Using Hamilton-type
equations, the dynamic equations of the scalar field variables (time
derivatives of $\chi$, $\phi$ and $\pi_{\chi}$, $\pi_{\phi}$) are obtained
together with the new terms added to the older dynamic equations.

We have realized also file-sequences in order to perform the reductional ADM
formalism for the canonical theory presented above. There is not an unique
method to realize this purpose. The procedure must be adapted to the
specific space-time model . The point is to guess a canonical transformation
(or a sequence of canonical transformations) which, after solving the
constraint equations and imposing specific coordinate conditions, generates
an action with one or two dynamic variables and a Hamiltonian generating the
time behavior of the system without constraints. These attempts to perform a
complete reductional formalism in each processed model can be a good way to
realize the connection with numerical relativity. We investigate the
possibility to generate FORTRAN lines for numerical solving of the
differential equations obtained after the canonical program outlined here is
performed.

When inflationary models are considered the above program lines are adapted
accordingly to the new dynamic and constraint equations (as we already
mentioned - see also \cite{6}) and, in addition we have new steps devoted to
the dynamical treatment of the scalar field. We shall use a different method
as a consequence of the fact that the scalar field has a unique component :

8) it is calculated the Dirac Hamiltonian from the super-Hamiltonian and
supermomentum obtained in steps 4 and 5 with formula (\ref{17}).

9) the dynamical equations for $\Phi $ and $\pi _\Phi $ are obtained using a
facility of EXCALC package, namely the variational derivative : 
\begin{verbatim}
derphi  :=   vardf(had,piphi);
derpiphi := - vardf(had,phi);
\end{verbatim}
where ''derphi'' and ''derpiphi'' denote the temporal derivatives of the
scalar field $\Phi $ and of his momentum $\pi _\Phi $ respectively. With
''had'' we have denoted the Dirac Hamiltonian already mentioned.

\section{About the concrete results}

In the above cited articles (\cite{4} -- \cite{7}) we have presented several
results obtained with space-time models processed by the procedures
presented here.

In order to verify the accuracy of our programs we approached some well
known models from the literature \cite{4},\cite{9}, \cite{16} - \cite{18}.

We have reobtained all the results reported in \cite{4}. Some of these
space-time models were processed in a generalized form in interaction with a
massless scalar field. Another well known case, the canonical treatment of
cylindrical gravitational waves \cite{18} (also coupled with a scalar field)
was one of the preferred testing models for us. The coincidence of our
results with the well-known results from \cite{18} was a good sign for
approaching other models, more sophisticated. Among these models is also the
space-time with $T^3$ three-dimensional subspace which we shall present
here. We give the form of the spatial metric tensor and of the momenta
canonically conjugate to the $g_{ij}$'s, the constraints (the
super-hamiltonian and the super-momenta) as well as the expressions of the
dynamic equations for all the canonical variables.

At the end of the list of specific results we have presented the skeleton of
the reductional procedure we have applied here : changing of variables,
canonical transformations, rescaling of variables, coordinate conditions
etc., and finally the reduced hamiltonian.

{\bf {Space-time model with a three sub-space in form of a 3-torus}} here
depicted as a 2-tori fibring of $S^1$ 
\begin{eqnarray} 
g_{ij} = \left ( \begin{array}{ccc}
                  e^{2(w-f)} & 0 & 0 \\
                  0 & e^{-2(w+f)}& 0 \\
                  0 & 0 & e^{2a}   \end{array} \right ) \nonumber
\end{eqnarray}
where $a = a(z,t)$, $w = w(z,t)$, $f = f(t)$ and $0 \leq x,y,z \leq 1$, and
the coframe is $w^1 = dx$, $w^2 = dy$, $w^3 = dz$. 
\begin{eqnarray} 
\pi^{ij} = \left ( \begin{array}{ccc}
          -\frac{1}{4} e^{-2(w-f)}(\pi_f-\pi_w) & 0 & 0 \\
            0 & -\frac{1}{4}e^{2(w+f)}(\pi_f+\pi_w) & 0 \\
            0 & 0 & \frac{1}{2} e^{-2a} \pi_a  \end{array}  \right )
\nonumber
\end{eqnarray}

The derivatives with respect to $z$ variable being denoted with "$\prime$"
we have 
\begin{eqnarray} 
{\cal{H}} = \frac{1}{8} e^{2f-a} \left [
            \pi_a^2 + 2\pi_a \pi_f + \pi_w^2 + 16 e^{-4f} w^{\prime 2}
                            \right ] +
          \frac{\pi_{\chi}^2}{4\sqrt{g}} + \sqrt{g} e^{2a} \chi^{\prime 2} = 0
\nonumber
\end{eqnarray}
\begin{eqnarray} 
{\cal{H}}^3 = e^{-2a} \left [ 
         \pi_a a^{\prime} - \pi_a^{\prime} + \pi_w w^{\prime} \right ] 
                  + e^{2a} \pi_{\chi} \chi^{\prime} = 0 
                  ~~~~~;~~~~~{\cal{H}}^1={\cal{H}}^2=0 \nonumber
\end{eqnarray}
\begin{eqnarray} 
\dot{f} = \frac{1}{4} e^{2f-a} N \pi_a \hbox{~~~;~~~}
\dot{w} = N^3 w^{\prime} + \frac{1}{4} e^{2f-a} N \pi_w \nonumber
\end{eqnarray}
\begin{eqnarray} 
\dot{a} = N^3 a^{\prime} + N^{3 \prime} + \frac{1}{4} e^{2f-a} N \pi_a 
                + \frac{1}{4}e^{2f-a} N \pi_f \hbox{~~~;~~~}
\dot{\chi} = N^3 \chi^{\prime} + \frac{1}{2} e^{2f-a} N \pi_{\chi} \nonumber
\end{eqnarray}
\begin{eqnarray}
\dot{\pi}_f = N^3 \pi_f^{\prime} +N^{3 \prime} \pi_f 
               - \frac{1}{4} N e^{2f-a} ( \pi_a^2 + 2\pi_a\pi_f + \pi_w^2 ) 
\nonumber\\
       -4 e^{-2f-a} (a^{\prime} N^{\prime} - N^{\prime \prime} - Nw^{\prime 2} )
        - \frac{1}{2} e^{-2f-a} N (e^{4f} \pi_{\chi}^2 - 4 \chi^{\prime 2} )
\nonumber
\end{eqnarray}
\begin{eqnarray} 
\dot{\pi}_{w}   =  \left ( N^3 \pi_w + e^{-a-2f} N w^{\prime} \right )^{\prime}
\nonumber
\end{eqnarray}
\begin{eqnarray} 
\dot{\pi}_a = N^3 \pi_a^{\prime} + N^{3 \prime} \pi_a +
              \frac{1}{8} e^{2f-a} (\pi_a^2 + 2\pi_a\pi_f + \pi_w^2)
             + 2 e^{-2f-a} N w^{\prime 2} \nonumber\\
             +\frac{1}{4} e^{-a-2f} N (e^{4f} \pi_{\chi}^2 + 4 \chi^{\prime 2})
\nonumber
\end{eqnarray}
\begin{eqnarray} 
\dot{\pi}_{\chi} = (N^3 \pi_{\chi} + 2 e^{-a-2f}N \chi^{\prime})^{\prime} 
\nonumber
\end{eqnarray}

- In order to perform a {\underline{reductional ADM formalism}} we have some
different possibilities, but the most cases lead to nonlocal theories. A
natural and very simple version is suggested by Misner, (in \cite{16} - but
not used in the cited article) is to choose $f$ as a time variable (the
transverse cross-sectional area of the universe is proportional to $e^{2f}$%
). Taking as coordinate variable $w$, solving the constraint equations in
terms of $\pi _w$ and $\pi _f$ and imposing the ADM coordinate conditions $%
T=f=t$ and $w=z$ we can obtain the action functional as : 
\begin{eqnarray} 
S = \int dt dz (\pi_a \dot{a} + \pi_{\chi} \dot{\chi} - H_{red} ) \nonumber
\end{eqnarray}
where the reduced time dependent hamiltonian $H_{red}$ is 
\begin{eqnarray}
H_{red} = - \pi_f = 
\frac{1}{2 \pi_a} (
             a^{\prime} \pi_a^2 + 2a^{\prime} \chi^{\prime} \pi_a \pi_{\chi}
           - 2 a^{\prime} \pi_a^{\prime} \pi_a + \chi^{\prime 2} \pi_{\chi}^2
           - 2 \chi^{\prime} \pi_a^{\prime} \pi_{\chi}\nonumber\\
            + \pi_a^{\prime 2}
           + \pi_a^2 + \pi_{\chi}^2 + 8 e^{-4T} \chi^{\prime 2} )
               \nonumber
\end{eqnarray}

Now, in the final lines, we shall mention only the results obtained after
processing the computer procedures outlined in the previous sections for
some space-time models, considered as initial data for the inflationary
universe described by (\ref{12}). The results reported in \cite{6} are only
in a brute form without any regarding of the physical significance.
Spherically-symmetric model (usual in the dynamic study of inflationary
universes), Bianchi I/Kantowski-Sachs models and also the above model with $%
T^3$ subspace were presented in details in \cite{6}.

\chapter{Using Maple in the study of the canonical formalism
of general relativity}
\chaptermark{Using Maple in the study ...}

\section{Introduction}

The use of computer facilities cam be an
important tool for teaching general relativity. We
have experienced several packages of procedures, (in REDUCE + EXCALC for
algebraic programming and in Mathematica for graphic visualizations) which
fulfill this purpose. In this chapter we shall present some new
procedures in MapleV using GrTensorII package (\cite{22}) adapted for the 
canonical version of the general relativity (in the so called ADM formalism 
based on the 3+1 split of spacetime). This formalism is widely used  in the 
last years as a major tool in numerical 
relativity for calculating violent processes as, for example the head-on 
collisions of black holes, massive stars or other astrophysical objects.
Thus we used these computer procedures in the process of teaching the canonical
formalism as an introductory part of a series of lectures on numerical
relativity for graduated students. We shall use the same notations and 
conventions already presented in the first chapter.
Obviously we used the programs in REDUCE presented there for producing our 
new procedures for Maple + GrTensorII
package, but because there are many specific features we shall present here
in some detail these procedures in the next section of the chapter. 
The last section of the chapter is dedicated to  the conclusions pointed
out by running the Maple procedures presented here and some future prospectives
on their usage toward the  numerical relativity.

\section{Maple + GrTensorII procedures}
\sectionmark{Maple + GrTensorII ...}

Here we shall describe briefly the structure and the main features of the Maple
procedures for the canonical formalism of the general relativity as described
in the previous chapter. Two major parts of the programs can be detected :
one before introducing the metric of the spacetime used (consisting in
several definitions of tensor objects which are common to all spacetimes)
and the second one,  having line-commands specific to each version.

The first part of the program starts after initalisation of the GrTensorII
package ({\bf grtw();}) and has mainly the next lines :
{\small
\begin{verbatim}
> grdef(`tr := pi{^i i}`);
> grdef(`ha0:=-sqrt(detg)*(Ricciscalar+
               (1/detg)*((1/2)*(tr)^2-pi{i j}*pi{ ^i ^j }))`);
> grdef(`ha{ ^i }:=-2*(pi{ ^i ^j ;j}-pi{ ^i ^j }*Chr{ p j ^p })`);
> grdef(`derge{ i j }:=2*N(x,t)*(detg)^(-1/2)*(pi{ i j } - 
                     (1/2)*g{ i j}*tr)+Ni{ i ;j } + Ni{ j  ;i }`);
> grdef(`Ndd{ ^m j }:= Nd{ ^m ;j }`); 
> grdef(`bum{ ^i ^j ^m}:=pi{ ^i ^j }*Ni{ ^m }`);
> grdef(`bla{ ^i ^j }:=bum{ ^i ^j ^m ;m }`);
> grdef(`derpi{ ^i ^j }:=
     -N(x,t)*(detg)^(1/2)*(R{ ^i ^j }-(1/2)*g{ ^i ^j }*Ricciscalar)+ 
      (1/2)*N(x,t)*(detg)^(-1/2)*g{ ^i ^j }*(pi{ ^k ^l   }*pi{ k l }-
      (1/2)*(tr)^2)-2*N(x,t)*(detg)^(-1/2)*(pi{ ^i ^m }*pi{ ^j m }-
      (1/2)*pi{ ^i ^j }*tr)+ (detg)^(1/2)*(Ndd{ ^i ^j }-g{ ^i ^j }*
       Ndd{ ^m m }) + bla{ ^i ^j } - Ni{ ^i ;m }*pi{ ^m ^j }-
       Ni{   ^j ;m }*pi{ ^m ^i }`);
\end{verbatim}
}
Here {\bf ha0} and {\bf ha\{~$\hat{}$ ~i ~\}} represents the superhamiltonian 
and the 
supermomentum as defined in eqs. (\ref{6}) and (\ref{7}) respectively and
{\bf tr} is the trace of momentum tensor density $\pi^{ij}$ - which will
be defined in the next lines of the program. Here {\bf N(x,t)}represents the
lapse function $N$. Also, {\bf derge\{ ~i ~j~\}} 
represents the time derivatives of the components of the metric tensor,
as defined in eq. (\ref{2}) and {\bf derpi\{~$\hat{}$~i~~$\hat{}$ ~j~\}} the
time derivatives of the components of the momentum tensor $\pi^{ij}$ as
defined in eq. (\ref{3}). 

The next line of the program is a specific GrTensorII command for loading
the spacetime metric. Here Maple loads a file (previously generated) for
introducing the components of the metric tensor as functions of the 
coordinates.
We also reproduced here the output of the Maple session showing the metric 
structure of the spacetime we introduced. 
\begin{verbatim}
> qload(`Cyl_din`);

                     Default spacetime = Cyl_din
                      For the Cyl_din spacetime:
                             Coordinates
                                x(up)
                             a
                           x   = [x, y, z]
                             Line element
     2                                    2
   ds  = exp(gamma(x, t) - psi(x, t))  d x
           2                     2                       2
  + R(x, t)  exp(-psi(x, t))  d y   + exp(psi(x, t))  d z
\end{verbatim}
As is obvious we introduced above the metric for a spacetime with cylindrical
symmetry, an example we used for teaching purposes being a well known example
in the literature (\cite{18}). In natural output this metric has the form :
\begin{equation}\label{cyl}
g_{ij} = \left ( \begin{array}{ccc}
e^{\gamma-\psi}&0&0\\
0&R^2 e^{-\psi}&0\\
0&0&e^{\psi}\end{array}\right )
\end{equation}
in cylindrical coordinates $x,y,z$ with
           $x \in [0 , \infty)$, 
           $y \in [0 , 2\pi)$,
           $z \in (-\infty , +\infty)$
where $R$, $\psi$ and $\gamma$ are functions of $x$ and $t$ only.

After the metric of the spacetime is established the next sequence of the 
programm just introduce the components of the momentum tensor $\pi^{ij}$ as
{\small
\begin{verbatim}
> grdef(`Nd{ ^ m } := [diff(N(x,t),x), 0, 0]`);
> grdef(`Ni{ ^i } := [N1(x,t), N2(x,t), N3(x,t)]`);
> grdef(`vi1{^i}:=[pig(x,t)*exp(psi(x,t)-gamma(x,t)),0,0]`);
> grdef(`vi3{^i}
       :=[0,0,exp(-psi(x,t))*(pig(x,t)+(1/2)*R(x,t)*pir(x,t)+
                                      pip(x,t))]`);
> grdef(`vi2{^i}:=[0,(2*R(x,t))^(-1)*pir(x,t)*exp(psi(x,t)),0]`);
> grdef(`pi{ ^i ^j } := 
          vi1{ ^i }*kdelta{^j $x}+vi2{ ^i }*kdelta{ ^j$y }+
          vi3{ ^i }*kdelta{^j $z}`);
> grcalc(pi(up,up));
> grdisplay(pi(up,up));
\end{verbatim}
}
Here {\bf Ni\{~$\hat{}$~i~\}} represents the shift vector $N^i$ and the other
objects ({\bf Nd}, {\bf vi1}, {\bf vi2} and {\bf vi3}) represent intermediate
vectors defined in order to introduce the momenum {\bf pi\{ ~$\hat{}$~i~
$\hat{}$~j~\}} having the form :
\begin{equation}\label{mom}
\pi^{ij}= \left ( \begin{array}{ccc}
\pi_{\gamma} e^{\psi-\gamma}& 0 & 0\\
0 & \frac{1}{2R}\pi_R e^{\psi} & 0 \\
0 & 0 & e^{-\psi}(\pi_{\gamma}+\frac{1}{2}R\pi_R + \pi_{\psi})\end{array}
\right )
\end{equation}
In the program we denoted $\pi_{\gamma}$, $\pi_R$ and 
$\pi_{\psi}$ with {\bf pig}, {\bf pir} and {\bf pip}, respectively. 
 The momentum components are introduced in order  that the dynamic 
part of the action of the theory be in  canonical form, that is : $\dot{g}_{ij}
\pi^{ij} = \pi_{\gamma}\dot{\gamma} + \pi_{\psi}\dot{\psi}+\pi_R\dot{R}$.
The next lines of the programm check if this condition is fullfiled :
{\small
\begin{verbatim}
> grdef(`de1{ i }:=[diff(grcomponent(g(dn,dn),[x,x]),t),0,0]`);
> grdef(`de2{ i }:=[0,diff(grcomponent(g(dn,dn),[y,y]),t),0]`);
> grdef(`de3{ i }:=[0,0,diff(grcomponent(g(dn,dn),[z,z]),t)]`);
> grdef(`ddgt({ i j }:=
       de1{ i }*kdelta{j $x}+de2{ i }*kdelta{ j$y }+
       de3{ i }*kdelta{ j $z}`);
> grcalc(ddgt(dn,dn));
> grdef(`act:=pi{ ^i ^j }*ddgt{ i j }`);
> grcalc(act); gralter(act,simplify); grdisplay(act);
\end{verbatim}
}
By inspecting this last output from the Maple worksheet, the user can decide
if it is necessary to redifine the components of the momentum tensor or to go
further. Here the components of the momentum tensor were calculated by hand
but,
of course a more experienced user can try to introduce here a
sequence of commands for automatic calculation of the momentum tensor
components using the above condition, through an intensive use of {\bf solve}
Maple command.

Now comes the must important part of the routine, dedicated to calculations
of different objects previously defined :
{\small
\begin{verbatim}
> grcalc(ha0); gralter(ha0,simplify);
> grdisplay(ha0);
> grcalc(ha(up)); gralter(ha(up),simplify);
> grdisplay(ha(up));
> grcalc(derge(dn,dn)); gralter(derge(dn,dn),simplify);
> grdisplay(derge(dn,dn));
> d1:=exp(-psi(x,t))*grcomponent(derge(dn,dn),[z,z])+exp(psi(x,t)-
      gamma(x,t))*grcomponent(derge(dn,dn),[x,x]);
> simplify(d1);
> d2:=(1/(2*R(x,t)))*exp(psi(x,t))*grcomponent(derge(dn,dn),[y,y])+
       (1/2)*R(x,t)*exp(-psi(x,t))*grcomponent(derge(dn,dn),[z,z]);
> simplify(d2);
> d3:=exp(-psi(x,t))*grcomponent(derge(dn,dn),[z,z]);
> simplify(d3);
> grcalc(derpi(up,up)); gralter(derpi(up,up),simplify);
> grdisplay(derpi(up,up));
> f1 := exp(gamma(x,t)-psi(x,t))*grcomponent(derpi(up,up),[x,x])-
                                                pig(x,t)*(d3-d1);
> simplify(f1);
>  f2:= 2*R(x,t)*exp(-psi(x,t))*grcomponent(derpi(up,up),[y,y])+
                             (1/R(x,t))*d2*pir(x,t)-pir(x,t)*d3;
> simplify(f2);
> f3 := exp(psi(x,t))*grcomponent(derpi(up,up),[z,z])+d3*(pig(x,t)+
        (1/2)*R(x,t)*pir(x,t)+pip(x,t))-f1-(1/2)*R(x,t)*f2-
                                                 (1/2)*pir(x,t)*d2;
> simplify(f3);
\end{verbatim}
}
This is a simple series of alternation of {\bf grcalc}, {\bf gralter} and
{\bf grdisplay} commands for obtainig the superhamiltonian, supermomentum
and the dynamic equations for the theory. {\bf d1 ... d3} and {\bf f1 ... f3}
are the time derivatives of the dynamic variables, $\dot{\gamma}$, $\dot{R}$,
$\dot{\psi}$ and $\dot{\pi}_{\gamma}$, $\dot{\pi}_{R}$, $\dot{\pi}_{\psi}$
respectively.
 Denoting with "$\prime$" the derivatives with respect to $r$ we display here
the results for the example used above (cylindrical gravitational waves)  :
{\small
\begin{eqnarray} 
    {\cal{H}}^0=e^{\frac{\psi-\gamma}{2}}
                  ( 2R^{\prime\prime} - R^{\prime}\gamma^{\prime} + 
                    \frac{1}{2}(\psi^{\prime})^{2}R -\pi_{\gamma}\pi_{R} +
                 \frac{1}{2R}(\pi_{\psi})^2 ) = 0 \hbox{~~~~~~~~~~~~~~~~~~} 
\nonumber
\end{eqnarray}
\begin{eqnarray} 
    {\cal{H}}^1={\cal{H}}^r=e^{\psi-\gamma}
                  ( -2\pi^{\prime}_{\gamma}+\gamma^{\prime}\pi_{\gamma}+
                    R^{\prime}\pi_R + \psi^{\prime}\pi_{\psi} ) = 0
                  \hbox{~~~~~~;~~~~~~}   {\cal{H}}^2 = {\cal{H}}^3 = 0 
\nonumber
\end{eqnarray}
\begin{eqnarray} 
    \dot{\gamma} = N^1\gamma^{\prime} +2N^{1\prime} - 
                               e^{ \frac{\psi-\gamma}{2}}N\pi_{R} 
\hbox{~~~~~;~~~~~}
    \dot{R} = N^1R^{\prime} - e^{\frac{\psi-\gamma}{2}}N\pi_{\gamma} 
\hbox{~~~~~~}
\nonumber
\end{eqnarray}
\begin{eqnarray} 
      \dot{\psi} = N^1\psi^{\prime} + \frac{1}{R}e^{\frac{\psi-\gamma}{2}}
N\pi_{\psi}
\hbox{~~~~~~;~~~~~~} 
\nonumber
\end{eqnarray}
\begin{eqnarray} 
     \dot{\pi}_{\gamma} = N^1 \pi_{\gamma}^{\prime} + N^{1 \prime} \pi_{\gamma}
                    -e^{ \frac{\psi - \gamma}{2} } ( R^{\prime} N^{\prime}
                 + \frac{1}{2} R^{\prime} \psi^{\prime} N 
                      - \frac{1}{4} \psi^{\prime 2} R N 
 + \frac{1}{2} N \pi_{\gamma}  \pi_R 
                - \frac{1}{4R} N \pi_{\psi}^2 )
\nonumber
\end{eqnarray}
\begin{eqnarray} 
      \dot{\pi}_{R} =  N^1 \pi_R^{\prime} + N^{1 \prime} \pi_R +
                   e^{\frac{\psi-\gamma}{2}} ( \gamma^{\prime} N^{\prime} 
             -2 N^{\prime \prime} - 2 N^{\prime} \psi^{\prime} 
                +\frac{1}{2} \gamma^{\prime} \psi^{\prime} N 
        - \psi^{\prime \prime} N - \psi^{\prime 2}
             +\frac{1}{2R} N \pi_{\psi}^2 ) 
\nonumber
\end{eqnarray}
\begin{eqnarray} 
         \dot{\pi}_{\psi} = N^1 \pi_{\psi}^{\prime} + N^{1 \prime} \pi_{\psi}
            + e^{ \frac{\psi -\gamma}{2} } ( R N^{\prime} \psi^{\prime}
             - R^{\prime \prime} N + \frac{1}{2} N R^{\prime} \gamma^{\prime}
                  + R^{\prime} \psi^{\prime} N 
             -\frac{1}{2} \gamma^{\prime} \psi^{\prime} N R
\nonumber
\end{eqnarray}
\begin{eqnarray} 
              +  \psi^{\prime \prime} R N + \frac{1}{4} \psi^{\prime 2} R N
              + \frac{1}{2} N \pi_R \pi_{\gamma} - \frac{1}{4R} N 
\pi_{\psi}^2 ) 
\nonumber
\end{eqnarray}
}
These are the well-known results reported in (\cite{18}) or (\cite{6}).

One of the important goals of the canonical formalism of the general relativity
(which constitutes the ``kernel'' of the ADM formalism) is the reductional
formalism. Here we obtain the true dynamical status of the theory, by reducing 
the number of the variables  through solving the constraint
equations. This formalism is applicable only to a restricted number of
space-time models, one of them being the above cylindrical gravitational
waves model. Unfortunately only a specific strategy can be used in every
model. Thus the next lines of our program must be rewritten specifically in
every case. Here, for teaching purposes we present our example of 
cylindrical gravitational wave space-time model. Of course we enccourage the 
student to apply his own strategy for other examples he dares to calculate.

In our example of cylindrical gravitational waves, the reductional
strategy as described in (\cite{18}) starts with the usual rescaling 
 of ${\cal{H}}$ and ${\cal{H}}^i$ to $\bar{{\cal{H}}}$ 
and $\bar{{\cal{H}}}^i$ by
\begin{eqnarray} 
  \bar{{\cal{H}}} = e^{ \frac{\gamma-\psi}{2} }{\cal{H}} \hbox{~~;~~}
          \bar{N} = e^{ \frac{\psi-\gamma}{2} }N
               \hbox{~~;~~}
  \bar{{\cal{H}}}^1 = e^{ \gamma-\psi }{\cal{H}}^1 \hbox{~~;~~}
          \bar{N}^1 = e^{ \psi-\gamma }N^1
\nonumber
\end{eqnarray}  
wich produce the next sequence of Maple+GrTensorII commands :
\begin{verbatim}
> grdef(`aha0:=sqrt(exp(gamma(x,t)-psi(x,t)))*ha0`);
> grdef(`aha{ ^j } := exp(gamma(x,t)-psi(x,t))*ha{ ^j }`);
> grdef(`an:=sqrt(exp(psi(x,t)-gamma(x,t)))*n(x,t)`);
> grdef(`ani{ ^i } := exp(psi(x,t)-gamma(x,t))*ni{ ^i }`);
\end{verbatim}
The canonical transformation to the new 
variables, including Kuchar's "extrinsic time", defined by :
\begin{eqnarray} 
    T = T(\infty) + \int_{\infty}^{r}(-\pi_{\gamma})dr
                \hbox{~~~,~~~}
    \Pi_{T} = -\gamma^{\prime} + [\ln{((R^{\prime})^2-(T^{\prime})^2})
]^{\prime}
\nonumber
\end{eqnarray}
\begin{eqnarray} 
    R = R
            \hbox{~~~,~~~}
    \Pi_{R} = \pi_{R} + 
       [\ln{(\frac{R^{\prime}+T^{\prime}}{R^{\prime}-T^{\prime}})}]^{\prime}
\nonumber
\end{eqnarray}
are introduced with :
\begin{verbatim}
> pig(x,t):=-diff(T(x,t),x);

> pir(x,t):=piR(x,t) - diff(ln((diff(R(x,t),x)+diff(T(x,t),x))/
                           (diff(R(x,t),x)-diff(T(x,t),x))),x);
\end{verbatim}
and specific substitutions in the dynamic objects of the theory :
{\small
\begin{verbatim}
> grmap(ha0, subs , diff(gamma(x,t),x)=
diff( ln( (diff(R(x,t),x))^2- (diff(T(x,t),x))^2 ),x)-piT(x,t),`x`);
> grcalc(ha0); gralter(ha0,simplify);
> grdisplay(ha0);
> grmap(ha(up), subs , diff(gamma(x,t),x)=diff( ln( (diff(R(x,t),x))^2- 
          (diff(T(x,t),x))^2 ),x)-piT(x,t),`x`);
> gralter(ha(up),simplify);
> grdisplay(ha(up));
> grcalc(aha0);
> grmap(aha0, subs , diff(gamma(x,t),x)=diff( ln( (diff(R(x,t),x))^2- 
                  (diff(T(x,t),x))^2 ),x)-piT(x,t),`x`);
> gralter(aha0,simplify,sqrt);
> grdisplay(aha0);
> grcalc(aha(up));
> grmap(aha(up), subs , diff(gamma(x,t),x)=diff( ln( (diff(R(x,t),x))^2- 
                  (diff(T(x,t),x))^2 ),x)-piT(x,t),`x`);
> gralter(aha(up),simplify);
> grdisplay(aha(up));
> grmap(act, subs , diff(gamma(x,t),x)=diff( ln( (diff(R(x,t),x))^2- 
                (diff(T(x,t),x))^2 ),x)-piT(x,t),`x`);
> grcalc(act); grdisplay(act);
\end{verbatim}
}
     Thus the action yields (modulo divergences) :
\begin{eqnarray} 
   S = 2 \pi \int_{-\infty}^{\infty} dt \int_{0}^{\infty} dr 
                  (\Pi_T \dot{T} + \Pi_R \dot{R} + \pi_{\psi} \dot{\psi}
                             + \pi_{\chi} \dot{\chi} -
                  \bar{N} \bar{{\cal{H}}} - \bar{N}_1 \bar{{\cal{H}}}^1 )
\nonumber
\end{eqnarray}
    where :
\begin{eqnarray} 
   \bar{{\cal{H}}} = R^{\prime} \Pi_T + T^{\prime} \Pi_R +
                \frac{1}{2}R^{-1} \pi_{\psi}^2 + \frac{1}{2}R \psi^{\prime 2}
                  +\frac{1}{4}R^{-1} \pi_{\chi}^2 + R \chi^{\prime 2}
\nonumber
\end{eqnarray}
\begin{eqnarray} 
   \bar{{\cal{H}}}^1 = T^{\prime} \Pi_T + R^{\prime} \Pi_R + 
                  \psi^{\prime} \pi_{\psi} +
                        \chi^{\prime} \pi_{\chi}
\nonumber
\end{eqnarray}
 
Solving the constraint equations  $\bar{{\cal{H}}}=0$ and 

$\bar{{\cal{H}}}^1=0$ for $\Pi_T$ and $\Pi_R$ and
imposing the coordinate conditions $T = t$ and $R = r$ we obtain finally :
\begin{eqnarray} 
    S = 2 \pi \int_{-\infty}^{+\infty} dT \int_{0}^{+\infty} dR 
              [\pi_{\psi} \psi_{,T} + \pi_{\chi} \chi_{,T} - 
                \frac{1}{2}( R^{-1} \pi_{\psi}^2 + R \psi_{,R}^2
                  +  R \pi_{\chi}^2 + R^{-1} \chi^{\prime 2}  )]
\nonumber
\end{eqnarray}
from the next sequence of programm lines :
\begin{verbatim}
> R(x,t):=x; T(x,t):=t; grdisplay(aha0);
> solve(grcomponent(aha0),piT(x,t));
> piT(x,t):= -1/2*(x^2*diff(psi(x,t),x)^2+pip(x,t)^2)/x;
> eval(piR(x,t));
> piR(x,t):=-diff(psi(x,t),x)*pip(x,t); piR(x,t);
> grdisplay(aha0); grdisplay(aha(up));
> piT(x,t);

                       2 /d           \2            2
                      x  |-- psi(x, t)|  + pip(x, t)
                         \dx          /
                - 1/2 -------------------------------
                                     x

> piR(x,t);

                       /d           \
                      -|-- psi(x, t)| pip(x, t)
                       \dx          /

> grcalc(act); grdisplay(act);
                      For the Cyl_din spacetime:

                                 act

                          /d           \
                    act = |-- psi(x, t)| pip(x, t)
                          \dt          /

> grdef(`Action:=act+piT(x,t)*diff(T(x,t),t)+piR(x,t)*diff(R(x,t),t)`);
> grcalc(Action);gralter(Action,factor,normal,sort,expand);
> grdisplay(Action);
                      For the Cyl_din spacetime:

                                Action


                   /d           \2   /d           \
  Action = - 1/2 x |-- psi(x, t)|  + |-- psi(x, t)| pip(x, t)
                   \dx          /    \dt          /

                        2
               pip(x, t)
         - 1/2 ----------
                   x

> grdef(`Ham:=piT(x,t)*diff(T(x,t),t)+piR(x,t)*diff(R(x,t),t)`);
> grcalc(Ham); gralter(Ham,expand);
> grdisplay(Ham);
                      For the Cyl_din spacetime:

                                 Ham
                                                         2
                          /d           \2       pip(x, t)
            Ham = - 1/2 x |-- psi(x, t)|  - 1/2 ----------
                          \dx          /            x

\end{verbatim}

\section{Conclusions. Further improuvements}

We used the programms presented above in the computer room with the students
from the graduate course on Numerical Relativity. The main purpose was to
introduce faster the elements of the canonical version of relativity with
the declared objective to skip the long and not very straitforward  hand 
calculations necessary to process an entire example of spacetime model. We
encouraged the students to try to modify the procedures in order to
compute new examples. 

The major conclusion is that this method is indeed usefull for an attractive
and fast teaching of the methods involved in the ADM formalism. On the
other hand we can use and modify these programs for obtaining the equations
necessary for the numerical relativity. In fact we intend to expand our
Maple worksheets for the case of axisymmetric model (used in the numerical
treatement of the head-on collision of black-holes). Of course, for numerical
solving of the dynamic equations obtained here we need more improuvements of
the codes for paralel computing and more sophisticated numerical methods.
But this will be the object of another article.

\chapter{Gravity, torsion, Dirac field and computer algebra}
\chaptermark{Gravity, torsion and ...}

\section{Introduction} 

\noindent In a series of recent published articles \cite{32}-\cite{33}
 we have presented some
routines and their applications written in REDUCE+EXCALC algebraic 
manipulation language for doing calculations in Dirac theory on curved
spacetimes.
Including the Dirac fields in gravitation theory requires lengthy
(or cumbersome)
calculations which appropriately could be solved by computer algebra
methods. 
Initially our main purpose was to develop a complete algebraic programming 
package for this purpose using only the REDUCE + EXCALC
platform. Partially this program was completed and the main results and 
applications were reported in our above cited articles \cite{32}-\cite{33}. 
But we are aware of the fact
that other very popular algebraic manipulations systems are on the market 
(like Mathematica or MAPLE) thus the area of people interested in algebraic
programming routines for Dirac equation should be much larger. In this 
perspective we developed similar programs and routines for MAPLE \cite{34}
 platform
using the package GRTensorII \cite{22}
(adapted for doing calculations in General 
Relativity). Because there is no portability between the two systems we were
forced to compose completely new routines, in fact using the same strategy
we used in REDUCE : first, the Pauli and Dirac matrices algebra (using only
the MAPLE environment) and then the construction of the Dirac equation on 
curved spacetimes were the capabilities of GRTensorII package is used. Because
the authors of GRTensorII offer also package versions  for Mathematica
we are sure that our  MAPLE routines can be easily adapted for Mathematica
with the result of highly increasing  the number of users of our product.

This chapter is organized as follows : the next section presents a short review
of the theory of Dirac fields on curved spacetime,  pointing out the main
notations and conventions we shall use. The next section
is devoted to a short overview of our routines and programs in REDUCE+EXCALC
previously described in great detail in \cite{32}. 
This section is necessary in view
of the fact that we used the same strategy for constructing our programs in
MAPLE as we pointed out above. Then the section nr. 4 contains a complete
description of our programs in MAPLE+GRTensorII. We also included here some
facts about the main differences (advantages and disadvantages) between the
two algebraic programming platforms (REDUCE and MAPLE). Section 5 is devoted
to the problem of including spacetimes with torsion in order to compute
Dirac equation using our MAPLE procedures. The last section of the article
includes a list of some of spacetimes examples we used in order to 
calculate the Dirac equation. Two of these examples are spacetimes with
torsion thus we pointed out the contribution of torsion components to the
Dirac field. Several applications of our programs (in REDUCE or in MAPLE) are
in view of our future research : searching for inertial effects in noninertial
systems of reference (partially presented in \cite{33} 
for a Schwarzschild metric
without torsion) or quantum effects (as in \cite{30}) in order to provide new
theoretical results  for  experimental gravity \cite{35}.

\section{Pauli and Dirac matrices algebra and Dirac equation on curved 
space-time}
\sectionmark{Pauli and Dirac ...}

The main problem is to solve algebraic expressions involving the 
Dirac matrices \cite{23},\cite{24}, \cite{26}. 
To this end it is convenient to construct explicitly  
these matrices as a direct product of several pairs among the Pauli matrices
$\sigma_i, i = 1,2,3$, and the $2 \times 2$ unit matrix. Thus all the 
calculations will be expressed in terms of the Pauli matrices and 
2-dimensional 
Pauli spinors. Consequently the  result will be obtained in a form 
which is suitable for physical interpretations. We shall consider the Pauli 
matrices as abstract objects with specific multiplication rules. 
Thus we work
with operators instead of their matrices in a spinor representation. However,
if one desires to see the result in the standard Dirac form with $\gamma$
matrices 
it will be sufficient to use a simple reconstruction procedure which will be
presented in a next section. 

We shall consider the Dirac formalism 
in the chiral form where the Dirac matrices are \cite{24} :
\begin{equation}\label{31}
\gamma^0 = \left [ \begin{array}{cc} 0 & 1 \\ 1 & 0 \end{array} \right ]
\hbox{~~,~~}
\gamma^i = \left [ \begin{array}{cc} 0 & \sigma_i \\ -\sigma_i & 0 \end{array}
\right ]\hbox{~~,~~}
\gamma^5 = \left [ \begin{array}{cc} -1 & 0 \\ 0 & 1 \end{array}
\right ]
\end{equation}
The Dirac spinor :
\begin{equation}\label{32}
\Psi = \left [ \begin{array}{cc} \varphi_l \\ \varphi_r \end{array} \right ] 
\in {\cal{H}}_D
\end{equation}
involves the Pauli spinors $\varphi_l$ and $\varphi_r$ which transform
according to the irreducible representations $(1/2,0)$ and $(0,1/2)$ of the
group SL(2,C). In this representation the left and right-handed Dirac 
spinors are
\begin{equation}\label{33}
\Psi_L = \frac{1-\gamma^5}{2} \Psi = \left [ 
          \begin{array}{c} \varphi_l \\ 0 \end{array} \right ] \hbox{~~,~~}
\Psi_R = \frac{1+\gamma^5}{2} \Psi = \left [ 
          \begin{array}{c} 0 \\ \varphi_r \end{array} \right ] 
\end{equation}
and, therefore, the Pauli spinors $\varphi_l$ and $\varphi_r$ will be the
left and the right-handed parts of the Dirac spinor. The SL(2,C) generators
are 
\begin{equation}\label{34}
S^{\mu\nu} = \frac{i}{4} \left [ \gamma^{\mu},\gamma^{\nu}\right ]
\end{equation}
It is clear \cite{23} that ${\cal{H}}_D = {\cal{H}} \otimes {\cal{H}}$ ( where 
${\cal{H}}$ is the two-dimensional space of Pauli spinors) and, therefore the
Dirac spinor can be written as:
\begin{equation}\label{35}
\Psi = \xi_1 \otimes \varphi_l + \xi_2 \otimes \varphi_r 
\hbox{~~with~~} \xi_1 = \left [ \begin{array}{c} 1 \\ 0 \end{array} \right ]
\hbox{~~and~~} \xi_2 = \left [ \begin{array}{c} 0 \\ 1 \end{array} \right ]
\end{equation}
while the $\gamma$-matrices and the SL(2,C) generators can be put in the form :
\begin{equation}\label{36}
\gamma^0 = \sigma^1 \otimes 1 \hbox{~~,~~} 
\gamma^k = i \sigma^2 \otimes \sigma^k \hbox{~~,~~}
\gamma^5 = - \sigma^3 \otimes 1 
\end{equation}
\begin{equation}\label{37}
S^{ij} = \frac{1}{2}\epsilon_{ijk} 1 \otimes \sigma^k \hbox{~~,~~} 
S^{0k} = - \frac{i}{2}\sigma^3 \otimes \sigma^k 
\end{equation}
These properties allow to reduce the Dirac algebra to that of the Pauli
matrices.

In order to introduce Dirac equation on curved space-time we have always used 
an anholonomic orthonormal frame because at any point of spacetime we need 
an {\it orthonormal reference frame}  to describe the spinor field 
as is already pointed before \cite{32}).
The Dirac equation in a general reference of frame, defined by an anholonomic 
tetrad field is \cite{27} :
\begin{equation}\label{38}
i\hbar\gamma^{\mu}D_{\mu}\Psi = mc \Psi 
\end{equation}
where the covariant Dirac derivative $D_{\mu}$ is 
\begin{equation}\label{39}
D_{\nu}=\partial_{\nu} +
  \frac{i}{2}S^{\rho \mu}\Gamma_{\nu \rho \mu} 
\end{equation}
and where $S^{\mu \nu}$ are the SL(2,C) generators (\ref{34}) and $\Gamma_{\nu \rho \mu}$
are the anholonomic components of the connection.

\section{Review of the REDUCE+EXCALC routines for calculating the Dirac equation}

\sectionmark{Review of REDUCE...}

We shall describe here those part of the program realizing the Pauli and Dirac 
matrix algebras. In the first lines of this sequence 
we introduce the operators and the non-commuting operators being 
useful throughout
the entire program. The Pauli matrices are represented using the operator 
{\bf p} with one argument. The Dirac matrices are denoted by {\bf gam} 
of one argument (an operator if we use only REDUCE, or for EXCALC package it 
will be a 0-form with one index) while the operator {\bf dirac} stands 
for the Dirac equation. The SL(2,C) generators  are denoted by the 
0-form {\bf s(a,b)}. The 
basic algebraic operation, the commutator ({\bf com}) and anticommutator 
({\bf acom}) are then defined here only for commuting (or anticommuting) 
only simple objects (``kernels''). For commuting more complex expressions, 
(in order to introduce some necessary commutation relations) a more complex 
operator is necessary to  introduce. 
Other objects, having a more or less local utilization in the program will 
be introduced with  declarations and statements at their specific appearance.

The main part of the program is the Pauli subroutine : 
\begin{verbatim}
        LET p(0)=1;                           
        LET p(2)*p(1)=-p(1)*p(2);             
        LET p(1)*p(2)=i*p(3);                 
        LET p(3)*p(1)=-p(1)*p(3);             
        LET p(1)*p(3)=-i*p(2);               
        LET p(3)*p(2)=-p(2)*p(3);            
        LET p(2)*p(3)=i*p(1);                 
        LET p(1)**2=1;                        
        LET p(2)**2=1;                        
        LET p(3)**2=1;                       
\end{verbatim}
The Pauli matrices, $\sigma_i$ appear as {\bf p(i)} while the $2 \times 2$
unity matrix is {\bf p(0)=1}. The properties of the Pauli matrices are
given by the above sequence of 10 lines.
The {\bf direct product} denoted by {\bf pd} operator has the properties 
introduced as :
\begin{verbatim}
for all a,b,c,u let pd(a,b)*pd(c,u)=pd(a*c,b*u);   
for all a,b let pd(a,b)**2=pd(a**2,b**2);          
for all a,b let pd(-a,b)=-pd(a,b);                 
for all a,b let pd(a,-b)=-pd(a,b);                 
for all a,b let pd(i*a,b)=i*pd(a,b);               
for all a,b let pd(a,i*b)=i*pd(a,b);               
for all a let pd(0,a)=0;                           
for all a let pd(a,0)=0;                           
let pd(1,1)=1;                                     
\end{verbatim}
\noindent  Some difficulties arise from the bilinearity of the direct 
product which
requires to identify all the scalars involved in the current calculations. 
This can be done only by using complicated procedures or special assignments.
For this reason we shall use a special definition of the direct product
({\bf pd}) which gives up the general bilinearity property. The operator 
{\bf pd} will depend on two Pauli matrices or on the Pauli matrices with 
factors $-1$ or $\pm i$. It is able to recognize only these numbers but this
is enough since the multiplication of the Pauli matrices has just the 
structure constants $\pm 1$ and $\pm i$ (we have $\sigma_i\sigma_j = 
\delta_{ij}+ i\epsilon_{ijk}\sigma_k$).

Thus by introducing the multiplication rules of the direct product it will 
be sufficient to give some instructions (see above) which represent the 
bilinearity defined only for the scalars $-1$ and $i$. The next two 
instructions represent the definition of ``$0$'' in the direct product space 
while the last one from the above sequence introduces the $4 \times 4$ unit 
matrix. 

The $\gamma$- matrices can be defined now with the help of our direct product;
also we added here the definition of the SL(2,C) generators from eq. 
(\ref{34}) :
\begin{verbatim}
gam(1):=i*pd(p(2),p(1));           
gam(2):=i*pd(p(2),p(2));            
gam(3):=i*pd(p(2),p(3));      %  Remember that gam's are
gam(0):=  pd(p(1),1);         %        0-forms !!!
gam5  := -pd(p(3),1);         %  instead of ``gam(5)''

s(a,b):= i*com(gam(a),gam(b))/4;                    
\end{verbatim}

 All the above program lines we have presented here can be 
used for the Dirac theory on the Minkowski space-time in an inertial system of 
reference. Here we shall first point the main differences which appear when 
one wants to run our procedures on some curved space-times or in a noninertial 
reference of frame. Some of these minor differences are already integrated 
in the lines presented in the precedent section.

First of all we must add, at the beginning of the program some EXCALC lines
containing the metric statement.  We have always used an anholonomic 
orthonormal frame because at any point of spacetime we need an {\it 
orthonormal reference frame} in order to describe the spinor field as is
already pointed before \cite{32}).

After the above metric statement we must add in the program
all the procedures described in the last section. Then we can
introduce some lines calculating the Dirac equation in this context.
 As a result we have used the next sequence 
of EXCALC lines :
\begin{verbatim}
pform {der(j),psi}=0; fdomain psi=psi(x,y,z,t);           
der(-j):= ee(-j)_|d psi + (i/2)*s(b,h)*cris(-j,-b,-h);    

operator derp0,derp1,derp2,derp3;                         
noncom derp0,derp1,derp2,derp3;                           

let @(psi,t)=derp0;                                       
let @(psi,x)=derp1;                    
let @(psi,y)=derp2;
let @(psi,z)=derp3;                                       

dirac := i*has*gam(j)*der(-j)-m*c;                        
ham:= -(gam(0)*(1/(ee(-0)_|d t))*dirac-i*has*derp0);      
\end{verbatim}
In defining the Dirac derivative {\bf der} we have introduced also an 
formal Dirac spinor ({\bf psi}) being a 0-form and depending on the 
variables imposed by the symmetry of the problem. It is just an intermediate
step (in fact a ``trick'') in order to obtain the partial derivative 
components as operators, because after calculating the components of the
covariant derivative ({\bf der(-j)}- see above) we have to replace
the partial derivatives of {\bf psi} with four non-commuting operators
{\bf derp0}, {\bf derp1} ... {\bf derp3}. 
The Dirac operator is thus defined as {\bf dirac}$:= (i \hbar \gamma^{\mu}
D_{\mu} - mc)\Psi$ and finally the Dirac Hamiltonian ({\bf ham}) is 
obtained from the canonical form of the Dirac equation :
$$i\hbar \frac{\partial \psi}{\partial t} = H \psi $$
which we shall use later, in the study of the nonrelativistic approximation
of the Dirac equation in noninertial reference frames. 

The results we have obtained after processing the program lines we have 
presented until now contains only the Pauli matrices and direct products of 
Pauli matrices. When one wish to have the final result in terms of the 
$\gamma$-matrices and SL(2,C) generators (and not in terms of direct products
of Pauli matrices) the procedure {\bf rec} will be used :
\begin{verbatim}
operator gama,gen;
noncom gama,gen;
PROCEDURE rec(a);
begin;
          ws1:=sub(pd(p(1),1)=gama(0),a);
          ws1:=sub(pd(p(2),1)=-i*gama(0)*gama(5),ws1);
          ws1:=sub(pd(p(3),1)=-gama(5),ws1);
          ws1:=sub(pd(1,p(1))=2*gen(2,3),ws1);
          ws1:=sub(pd(1,p(2))=2*gen(3,1),ws1);
          ws1:=sub(pd(1,p(3))=2*gen(1,2),ws1);
     for k:=1:3 do 
        <<ws1:=sub(pd(p(2),p(k))=-i*gama(k),ws1);
          ws1:=sub(pd(p(1),p(k))=gama(k)*gama(5),ws1);
          ws1:=sub(pd(p(3),p(k))=2*i*gen(0,k),ws1)>>;
   return ws1;
end;
\end{verbatim}
\noindent This is an operator depending on an expression involving matrices
({\bf a}) which reconstructs the $\gamma$-matrices and the SL(2,C) generators 
from the direct products of Pauli matrices according to eq. \ref{36} and 
\ref{37}. 

As a very important remark we must point out  that the new introduced 
operators {\bf gama} and {\bf gen} does not represent a complete algebra.
They are introduced in order to have the result in a comprehensible
form. Thus this form of the result cannot be used in further computations. 
Only the results obtained
before processing the {\bf rec} procedure can be used, in order to benefit
of the complete Pauli and Dirac matrices algebra. We  have used this {\bf rec}
procedure only for pointing out our results in a more comprehensible form.

\section{MAPLE+GRTensorII procedures for calculating the Dirac equation}
\sectionmark{MAPLE+GrTensorII procedure...}

Here we shall present, in details, our procedures in MAPLE+GRTensorII for
calculating the Dirac equations, pointing out the main differences between
MAPLE and REDUCE programming in obtaining the same results. The first major
problem appears in MAPLE when one try to introduce the Pauli and Dirac matrices
algebra. In MAPLE this will be a difficult task because the ordinary product
(assigned in MAPLE with ``$*$'') of operators is automatically commutative,
associative, linear, etc. like an ordinary scalar product - in REDUCE these
properties are active only if the operators are declared previously as having
such properties. Thus we have to define two special product operators :
for Pauli matrices $\sigma_{\alpha}, \alpha=0,1,2,3$ 
(assigned in our procedures with {\bf pr}) 
and for the direct product of Pauli matrices (assigned here also with 
the operator {\bf pd(``,``)}) which is assigned with ``{\bf \&p}''. As a 
consequence we have to introduce long lists with their properties as :
\begin{verbatim}
> define(sigma,sigma(0)=1);

> define(pr,pr(1,1)=1,pr(1,sigma(1))=sigma(1),pr(1,sigma(2))=
sigma(2),pr(1,sigma(3))=sigma(3),pr(sigma(1),1)=sigma(1),
pr(sigma(2),1)=sigma(2),pr(sigma(3),1)=sigma(3),...
pr(sigma(1),sigma(1))=1,pr(sigma(2),sigma(2))=1,...

> define(pd,pd(0,a::algebraic)=0,pd(a::algebraic,0)=0,pd(1,1)=1,
pd(I*a::algebraic,b::algebraic)=I*pd(a,b),
pd(-I*a::algebraic,b::algebraic)=-I*pd(a,b),
pd(a::algebraic,I*b::algebraic)=I*pd(a,b),
pd(a::algebraic,-I*b::algebraic)=-I*pd(a,b),
pd(-a::algebraic,b::algebraic)=-pd(a,b),
pd(a::algebraic,-b::algebraic)=-pd(a,b));

> define(`&p`,`&p`(-a::algebraic,-b::algebraic)=`&p`(a,b),...
\end{verbatim}
We dropped the complete list of the properties of the special products 
{\bf pr} and {\bf \&p}, being very
long. Of course the reader may ask why is 
not much simpler to declare, as an example the $\&p$ as being linear (or 
multilinear) ? Because in this case the operator does not act properly, the 
linearity property  picking out from the operator all the terms, being or not
Pauli matrices or direct products {\bf pd} of Pauli matrices. Thus is 
necessary to forget the linearity and to introduce, as separate properties 
all the possible situations to appear in the calculus. The result is that the
program become very large with a corresponding waste of RAM memory and speed
of running. This will be the main disadvantage of MAPLE version of our 
program in comparison with the short (and, why not, elegant) REDUCE procedures.
Of course, in a more compact version of our programs, we defined MAPLE routines
with these operators, and the user need only to load at the beginning of MAPLE 
session these routines, but there is no significative economy of memory and
running time. 

The next step is to define Dirac $\gamma$-matrices and a special commutator
(with ${\bf \&p}$) :
\begin{verbatim}
> define(gam,
gam(1)=I*pd(sigma(2),sigma(1)),
gam(2)=I*pd(sigma(2),sigma(2)),
gam(3)=I*pd(sigma(2),sigma(3)),
gam(0)=pd(sigma(1),1),gam(5)=-pd(sigma(3),1));

> define(comu,comu(a::algebraic,b::algebraic)=a &p b - b &p a);
\end{verbatim}

The next program-lines are in GRTensorII environment. For this is necessary to
load previously the GRTensorII package and then to load the corresponding
metric (with {\bf qload(...)} command. It follows then :
\begin{verbatim}
> grdef(`SS{ ^a ^b }`);
> grcalc(SS(up,up)):
> (I/4)*comu(gam(0),gam(0));
> (I/4)*comu(gam(1),gam(0));
      .
      .
      .
> (I/4)*comu(gam(3),gam(3));
> grdisplay(SS(up,up));

> grcalc(Chr(dn,dn,dn));grdisplay(Chr(dn,dn,dn));
> grcalc(Chr(bdn,bdn,bdn));grdisplay(Chr(bdn,bdn,bdn));
\end{verbatim}
These are a sequence of commands in GRTensorII for defining the SL(2,C) 
generators $S^{ij}$ (as the tensor {\bf SS\{ \^{ }a \^{ }b\}}) using formula 
(\ref{34})
and for the calculus of Christoffel symbols in an orthonormal frame base 
({\bf Chr(bdn,bdn)}). Here is active one of the main advantages of 
MAPLE+GRTensorII
platform, namely the possibility of the calculus of the tensor components 
both in a general reference frame or in an anholonomic orthonormal frame 
which is vital for our purpose of construction the  Dirac equation. 

Next we have to define, as two vectors  the Dirac-$\gamma$ matrices (assigned 
as the contravariant vector {\bf ga\{ \^{ }a \}} and the derivatives of the wave
function $\psi$ (assigned as the covariant vector {\bf Psid\{ a \}} in order
to use the facilities of GRTensorII to manipulate with indices :
\begin{verbatim}
> grdef(`ga{ ^a }:=[gam(0),gam(1),gam(2),gam(3)]`);
> grdisplay(ga(up));

> grdef(`Psid{ a }:=[diff(psi(x,t),t),diff(psi(x,t),x),
                      diff(psi(x,t),y),diff(psi(x,t),z)]`);
> grcalc(Psid(dn));grdisplay(Psid(dn));
> grcalc(Psid(bdn));grdisplay(Psid(bdn));
\end{verbatim}

The next step defines a term which will be the term $\frac{i}{2}
S^{\rho\mu}\Gamma_{\nu\rho\mu}$ from equation \ref{39} :
\begin{verbatim}
> grdef(`de{ i }:=(I/2)*SS{ ^a ^b }*Chr{ (i) (a) (b) }`); 
> grcalc(de(dn));grdisplay(de(dn));
\end{verbatim}
Observing that the components of {\bf de\{ i \}} are polynoms containing direct
products {\bf pd(...)} of Pauli matrices and the fact that the product
between $\gamma^{\nu}$ and  $\frac{i}{2}S^{\rho\mu}\Gamma_{\nu\rho\mu}$ 
from equation \ref{38} is, in fact the special product $\&p$ we have to 
obtain the term
$\gamma^{\nu} S^{\rho\mu}\Gamma_{\nu\rho\mu}$ 
(denoted below with the operator {\bf dd}) 
by a special MAPLE sequence which in fact split the components of 
{\bf de\{ i \}} in monomial terms and then execute the corresponding $\&p$ product, finally reconstructing the {\bf dd} operator :
\begin{verbatim}
 a0:=expand(grcomponent(de(dn),[t]));a00:=0;
> u0:=whattype(a0);u0; nops(a0);
> if u0=`+` then  for i from 1 to nops(a0) do 
     a00:=a00+I*h*grcomponent(ga(up),[t]) &p op(i,a0) od else 
     a00:=I*h*grcomponent(ga(up),[t]) &p a0 fi; a00;

> a1:=expand(grcomponent(de(dn),[x]));a11:=0;
> u1:=whattype(a1);u1;
> nops(a1);
> if u1=`+` then  for i from 1 to nops(a1) do 
     a11:=a11+I*h*grcomponent(ga(up),[x]) &p op(i,a1) od else 
     a11:=I*h*grcomponent(ga(up),[x]) &p a1 fi;a11;
      .
      .
      .
> grdef(`dd`); grcalc(dd);
> a00+a11+a22+a33;
> grdisplay(dd);
\end{verbatim}
Finally the Dirac equation is obtained as :
\begin{verbatim}
> grdef(`dirac:=
   I*h*ga{ ^l }*Psid{ (l) } + dd*psi(x,t) - m*c*psi(x,t)`);
> grcalc(dirac);
> grdisplay(dirac);
\end{verbatim}
In order to obtain the Dirac equation in a more comprehensible form we have the
next sequence of MAPLE commands (similar to the reconstruction {\bf rec} 
procedure from the REDUCE program :
{\small
\begin{verbatim}
> define(`gen`);
> define(`gama`);
> grmap(dirac,subs,pd(sigma(1),1)=gama(0),pd(sigma(2),1)=-I*gama(0)*
gama(5),pd(sigma(3),1)=-gama(5),pd(1,sigma(1))=2*gen(2,3),pd(1,
sigma(2))=2*gen(3,1),pd(1,sigma(3))=2*gen(1,2),pd(sigma(2),sigma(1))=
-I*gama(1),pd(sigma(2),sigma(2))=-I*gama(2),pd(sigma(2),sigma(3))=
-I*gama(3),pd(sigma(1),sigma(1))=gama(1)*gama(5),pd(sigma(1),sigma(2))=
gama(2)*gama(5),pd(sigma(1),sigma(3))=gama(3)*gama(5),pd(sigma(3),
sigma(1))=2*I*gen(0,1),pd(sigma(3),sigma(2))=2*I*gen(0,2),pd(sigma(3),
sigma(3))=2*I*gen(0,3),`x`);
> grdisplay(dirac);
\end{verbatim}
}
where, of course, as in REDUCE version, the operators {\bf gen} and {\bf gama}
does not represent a complete algebra.

\section{Dirac equation on spacetimes with torsion and computer algebra}
\sectionmark{Dirac eq. ... with torsion ...}

We shall present here how we adapted the already presented in the previous 
section our MAPLE+GRTensorII programs in order to calculate the Dirac equation
on space-times with torsion. 

The geometrical frame for General Relativity is a Riemannian  
space--time but one very promising generalization is the Riemann--Cartan 
geometry  which (i) is the most natural generalization of a Riemannian 
geometry by allowing a non--symmetric metric--compatible connection, (ii) 
treats spin on the same level as mass as it is indicated by the group  
theoretical analysis of the Poincar\'e group, and (iii) arises in most  
gauge theoretical approaches to General Relativity, as e.g. in the  
Poincar\'e--gauge theory  or supergravity \cite{26},\cite{27}.
However, till now there is no experimental evidence for torsion. 
On the other hand, from the lack of effects which may be due to  
torsion one can calculate estimates on the maximal strength of the  
torsion fields \cite{30}. 
In this aspect we think that is possible, using computer
algebra facilities to approach new theoretical aspects on matter fields 
(for example the Dirac field) behavior on spacetimes with torsion in order
to point out new gravitational effects and experiments at microscopic level.

A metric compatible connection components in a Riemann-Cartan theory is 
related to the torsion components by (see \cite{26} - eq. (1.18))
\begin{equation}\label{tor1}
\Gamma_{\alpha \beta \gamma} = \tilde{\Gamma}_{\alpha \beta \gamma} -
\frac{1}{2}\left[ (C_{\alpha \beta \gamma} - C_{\beta \gamma \alpha} +
C_{\gamma \alpha \beta}) -
(T_{\alpha \beta \gamma} -T_{\beta \gamma \alpha} + T_{\gamma \alpha \beta})
\right]
\end{equation}
where $\tilde{\Gamma}_{\alpha \beta \gamma}$ are the components of the 
riemannian connection, $C_{\alpha \beta\gamma}$ is the object of anholonomicity
and $T_{\alpha \beta \gamma}$ are the components of the torsion.
The idea is to replace the connection components from
the covariant derivative appearing in eqs. (\ref{38}-\ref{39}) 
with the above ones,
of course after calculating then in an orthonormal anholonomic reference frame
suitable for calculation of the Dirac equation. Thus the sequence for 
calculating the {\bf de\{dn\}} operator (see above) should be replaced with
{\small
\begin{verbatim}
> grdef(`de{ i }:=(I/2)*SS{ ^a ^b }*(CHR{ (i) (a) (b) }+
                   (1/2)*(tor{ (i) (a) (b) } -tor{ (a) (b) (i) } 
                                    + tor{ (b) (i) (a) })))`);
> grcalc(de(dn)); grdisplay(de(dn));
\end{verbatim}
}
where the new connection components $\bf CHR{a,b,c}$ are now defined by the
sequence
{\small
\begin{verbatim}
> grdef(`ee{ a ^b }:= w1{ ^b }*kdelta{ a $x } 
                       + w2{ ^b }*kdelta{ a $y } + 
          w3{ ^b }*kdelta{ a $z } + w4{ ^b }*kdelta{ a $t }`); 
> grcalc(ee(dn,up));
> grdisplay(ee(dn,up));
> grdef(`CC{ a b c }:=2*ee{ a ^i }*ee{ b ^ j }*ee{ c [j ,i] }`);
> grcalc(CC(dn,dn,dn)); grdisplay(CC(dn,dn,dn));
> grdef(`CHR{ (a) (b) (c) } := Chr{ (a) (b) (c) } - 
                                    (1/2)*(CC{ a b c } - 
                                CC{ b c a } + CC{c a b })`);
> grcalc(CHR(bdn,bdn,bdn)); grdisplay(CHR(bdn,bdn,bdn));
\end{verbatim}
}
and the rest of the routines are unchanged. The only problem remains now to
introduce in an adequate way the components of the torsion tensor. We used
a suggestion from \cite{26} 
pointing that we can assume that the 2-form $T^{\alpha}$
associated to the torsion tensor should have the same pattern as the $d\theta^{
\alpha}$'s where $\theta^{\alpha}$ is the orthonormal coframe, who's
components are denoted in GRTensorII with {\bf w1\{dn\} ...w4\{dn\}}. Thus
this operation it is  possible only after we introduced the metric (with
{\bf qload} command). Calculating then the derivatives of the orthonormal frame
vector basis components we can introduce the torsion components by 
inspecting carefully these derivatives. Here is an example of how this is
possible in MAPLE+GRTensorII using one of the metric examples presented in
the next section:
\begin{verbatim}
> grcalc(w1(dn,pdn));grcalc(w2(dn,pdn));
> grcalc(w3(dn,pdn));grcalc(w4(dn,pdn));
> grdisplay(w1(dn,pdn)); grdisplay(w2(dn,pdn));
> grdisplay(w3(dn,pdn)); grdisplay(w4(dn,pdn));
> grcalc(w1(bdn,pbdn));grcalc(w2(bdn,pbdn));
> grcalc(w3(bdn,pbdn));grcalc(w4(bdn,pbdn));
> grdisplay(w1(bdn,pbdn));grdisplay(w2(bdn,pbdn));
> grdisplay(w3(bdn,pbdn));grdisplay(w4(bdn,pbdn));
> grdef(`tor{ ^a b c }:=w1{ b ,c }*kdelta{ ^a $x } + 
                        w2{ b ,c }*kdelta{ ^a $y } + 
                        w3{ b ,c }*kdelta{ ^a $z } + 
                        w4{ b ,c }*kdelta{ ^a $t }`);
> grcalc(tor(up,dn,dn));grdisplay(tor(up,dn,dn));
> grmap(tor(up,dn,dn),subs,f(x)=v4(x),h(x)=v3(x),g(x)=v2(x),`x`);
> grdisplay(tor(up,dn,dn));
> grcalc(tor(bup,bdn,bdn));
> grdisplay(tor(bup,bdn,bdn));
\end{verbatim}
The reader can observe that we first assigned the components of the torsion 
tensor (here denoted with {\bf tor\{up,dn,dn\}}) then after displaying his
components we ca decide to substitute new functions describing the torsion
instead of the functions describing the metric. Of course finally we calculate
the components of the torsion in an orthonormal anholonomic reference frame
({\bf tor\{bup,bdn,bd\}}).

\section{New results}

This section is devoted to a list of more recent results we obtained by
running our procedures in MAPLE+GRTensorII already described in the previous
sections. First we tested our programs by re-obtaining the form of the Dirac
equations in several spacetime metrics we already obtained with REDUCE+EXCALC
procedures and reported in our articles \cite{32}-\cite{33}. 
The concordance of these
results with the previous ones was a good sign for us to proceed with more
complicated and new examples, including ones with torsion. Here we shall
present some of these examples. \\

\noindent {\bf 1.} {\it Conformally static metric with  $\Phi$ and $\Sigma$
constant} where the line element is :
\begin{eqnarray}
ds^2 = e^{2\Phi t+2\Sigma}a(r)^2 dr^2 +e^{2\Phi t+2 \Sigma}r^2 d\theta^2 
+e^{2\Phi t+2 \Sigma}r^2~sin(\theta)^2 d\phi^2 -
e^{2\Phi t+2\Sigma}b(r)^2 dt^2 \nonumber
\end{eqnarray}
Thus the Dirac equation becomes :
{\small
\begin{eqnarray}
i\hbar e^{-\Phi t - \Sigma}\left[ \gamma^1 \left(\frac{1}{a(r)}\frac{\partial}{
\partial t} -\frac{1}{2a(r)b(r)}b^{\prime}(r) -\frac{1}{a(r)r}\right) 
+\gamma^2
\left(\frac{1}{r}
\frac{\partial}{\partial r} + \frac{1}{r~sin(\theta)}\frac{\partial}{\partial 
\theta} \right . \right .\nonumber\\
\left . \left .
- \frac{1}{2r}~cotg(\theta)\right) 
+ \frac{1}{b(r)} \gamma^0 \left( \frac{3}{2}
- \frac{\partial}{\partial \phi}\right)\right]\psi - mc\psi=0\nonumber
\end{eqnarray}
}
where $b^{\prime}(r)$ is the derivative $\partial b(r)/\partial r$.\\
 
\noindent {\bf 2.} {\it Taub-NUT spacetime} having the line element as :
\begin{eqnarray}
ds^2 = -4 l^2 U(t)dy^2 -8l^2 U(t)~cos(\theta)dy d\phi -(t^2+l^2)d\theta^2 
~~~~~~~~~~~~~\nonumber\\
~~~~~~~~~~~~~+(-4l^2U(t)~cos(\theta)^2-(t^2+l^2)~sin(\theta)^2)d\phi^2 
+ \frac{1}{U(t)}dt^2\nonumber
\end{eqnarray}
the coordinates being $(y,\theta,phi,t)$. We obtained the Dirac equation as :
{\small
\begin{eqnarray}
\frac{i\hbar}{t^2+l^2}\left[-\gamma^0\left(\frac{t^2+l^2}{4\sqrt{U(t)}}
U^{\prime}(t) + \sqrt{U(t)}\left(1 +\frac{\partial}{\partial t}\right)\right)-
cotg(\theta)\sqrt{t^2+l^2}\gamma^2\right]\psi(t) \nonumber\\
- mc\psi(t)=0\nonumber
\end{eqnarray}
~~~~~~~~~\\
}
\noindent {\bf 3.} {\it G\" odel spacetime}, having the line element as :
\begin{eqnarray}
ds^2 = -a^2 dx^2 +\frac{1}{2}a^2 e^{2x}dy^2 +2a^2 e^{x}cdy dt -a^2 dz^2 +a^2 
c^2dt^2
\nonumber
\end{eqnarray}
in coordinates $(x,y,z,t)$. Here the Dirac equation is simply :
{\small
\begin{eqnarray}
i\hbar\frac{1}{a}\left [ - \gamma^1\left (\frac{1}{2}+\frac{\partial}{\partial
x}\right)+\sqrt{2}\gamma^2\left(e^{-x}\frac{\partial}{\partial y} -\frac{1}{c}
\frac{\partial}{\partial t}\right)-\gamma^3\frac{\partial}{\partial z}+
\frac{1}{c}\frac{\partial}{\partial t}\right]\psi(x,y,z,t) \nonumber\\
- mc\psi(x,y,z,t)=0
\nonumber
\end{eqnarray}
}
\noindent {\bf 4.} {\it McCrea static spacetime} \cite{26} with {\bf torsion} 
having the line element as
\begin{eqnarray}
ds^2 = -e^{2f(x)}dt^2 + dx^2 + e^{2g(x)}dy^2 + e^{2h(x)}dz^2\nonumber
\end{eqnarray}
in $(x,y,z,t)$ coordinates. If the coordinate lines of $y$ are closed with
$0\leq y < 2\pi$ and $-\infty < z < \infty$, $ 0 < x < \infty$, the spacetime
is cylindrically with $y$ as the angular, $x$ the cylindrical radial and $z$ 
the longitudinal coordinate. If $-\infty , x,y,z < \infty$ the symmetry is pseudo-planar. In \cite{26} 
McCrea considers the simplest solution of Einstein equations
with cosmological constant as
\begin{equation}\label{Mc}
f=h=h=qx/3
\end{equation}
and the cosmological constant turns to be $q^2/3$. We shall first consider
the general case specializing the results at the final step of the program 
to the above particular solution. Running our MAPLE+GRTensorII procedures
first we obtain the torsion tensor component as :
\begin{eqnarray}
{T}^y_{yx} = \frac{\partial v2(x)}{\partial x}e^{v2(x)}\hbox{~;~~}
{T}^z_{zx} = \frac{\partial v3(x)}{\partial x}e^{v3(x)}\hbox{~;~~}
{T}^t_{tx} = \frac{\partial v4(x)}{\partial x}e^{v4(x)}\nonumber
\end{eqnarray}
the rest of the components being zero. This time we have obtained the Dirac 
equation, depending also on the components of the torsion tensor as :
\begin{eqnarray}
i\hbar \frac{1}{2}\gamma^1\left ( e^{v2(x)}\frac{\partial v2(x)}{\partial x}+
e^{v3(x)}\frac{\partial v3(x)}{\partial x}+e^{v4(x)}\frac{\partial v4(x)}
{\partial x}-\frac{\partial f(x)}{\partial x}-\frac{\partial g(x)}{\partial x}
\right. \nonumber\\
\left.
-\frac{\partial h(x)}{\partial x} +2\frac{\partial}{\partial x}\right)\psi(x)-
mc\psi(x)=0\nonumber
\end{eqnarray}
Of course when we specialize to the particular solution proposed by McCrea in
\cite{6} we have to assign the form of metric functions as in (\ref{Mc}) and we can
then take the torsion functions as
\begin{eqnarray}
v2=v3=v4=v(x)\nonumber
\end{eqnarray}
and the Dirac equation becomes
\begin{eqnarray}
i\hbar\gamma^1\left(\frac{3}{2}\frac{\partial v(x)}{\partial x}e^{v(x)}-
\frac{1}{2}q + \frac{\partial}{\partial x}\right)\psi(x) -mc\psi(x)=0\nonumber
\end{eqnarray}

\noindent {\bf 5} {\it Schwarzschild metric} with {\bf torsion}. This example
it is interesting in view of recent investigations on the contribution of
torsion in gravity experiments using atomic interferometry \cite{30}. 
Here we used
the Schwarzschild metric having the line element written as
\begin{eqnarray}
ds^2 = e^{\lambda(r)}dr^2 `+r^2d\theta^2 +r^2~sin(\theta)^2 d\phi^2 -
e^{\nu(r)}dt^2 \nonumber
\end{eqnarray}
Here we preferred to specialize the form of $\lambda(r)$ and $\nu(r)$ functions
as
\begin{eqnarray}
\nu(r) = 1-\frac{2m}{r}=\frac{1}{\lambda(r)}\nonumber
\end{eqnarray}
(here we have $G=c=1$) later after obtaining the form of the Dirac equation
in term of $\lambda(r)$ and $\nu(r)$ functions.

Using the same ``trick'' as in the previous example, we choose the components
of the torsion tensor as
\begin{eqnarray}
{T}^r_{r r} = \frac{1}{2}\frac{\partial f1(r)}{\partial r}e^{1/2 f1(r)}
\hbox{~;~~}
{T}^{\theta}_{\theta r}=1
\hbox{~;~~}
{T}^{\phi}_{\phi r}=sin(\theta)\nonumber
\end{eqnarray}
\begin{eqnarray}
{T}^t_{t r} = \frac{1}{2}\frac{\partial f2(r)}{\partial r}e^{1/2 f2(r)}
\hbox{~;~~}
{T}^{\phi}_{\phi \theta} = r~cos(\theta)
\nonumber
\end{eqnarray}
the rest of the components being zero. Running away our procedures we obtain
the Dirac equation containing terms with torsion tensor components as :
\begin{eqnarray}
i\hbar\left[ \frac{1}{4}e^{-\lambda(r)/2}\gamma^1
\left(\frac{\partial f2(r)}{\partial 
r}e^{f2(r)/2} - \frac{\partial \nu(r)}{\partial r} -\frac{4}{r} +2(1+sin(
\theta))+4\frac{\partial}{\partial r}\right) +\right .\nonumber\\
\left .\frac{1}{2}\gamma^2(cos(\theta)-
cotg(\theta))\right]\psi(r) - m\psi(r)=0\nonumber
\end{eqnarray}

\end{document}